\documentclass{iopjournal}

\usepackage[utf8]{inputenc} 
\usepackage[T2A]{fontenc}
\usepackage{hyperref}
\usepackage{amssymb,amsmath}
\usepackage{multirow}
\usepackage{caption}
\usepackage{graphicx}

\begin{document}

\articletype{Paper} 

\title{The way to the Big Bang}

\author{V. A. Berezin$^1$\orcid{0000-0002-2695-174X}, I. D. Ivanova$^{2,*}$\orcid{0000-0001-7570-0624} and A. E. Kuprina$^3$}

\affil{$^1$ Theory Physics Department, Institute for Nuclear Research of the Russian Academy of Sciences, Moscow, Russia}

\affil{$^2$ Physics Department, Ben-Gurion University of the Negev, Beer-Sheva, Israel}

\affil{$^3$ Institute of Education, National Research University "Higher School of Economics", Moscow, Russia}

\affil{$^*$Author to whom any correspondence should be addressed.}

\email{pc\textunderscore mouse@mail.ru}


\begin{abstract}
We propose conformal invariance as a fundamental symmetry governing cosmological particle creation from vacuum fluctuations, employing a phenomenological approach with an ideal fluid action to address the long-standing back-reaction problem. We demonstrate that particle production cannot emerge from classical vacua but must originate from a quantum vacuum at zero scale factor, with the transition surface constituting a light-like rather than space-like hypersurface. This implies that particles are created on the light cone and remain causally connected, with their apparent simultaneity being illusory. Our model requires an open Universe ($k=0, -1$) and reconceptualizes the Big Bang as a detonation wave propagating through quantum vacuum at the speed of light.
\end{abstract}
\section{Introduction} 

Additional symmetry increases the probability of the Universe emerging from "nothing" \cite{Vilenkin}. We propose that conformal invariance is the fundamental symmetry we have been seeking. This idea has support from, among others, Roger Penrose \cite{Penrose} and Gerard 't Hooft \cite{Hooft}.

The conformally invariant gravitational Lagrangian contains terms quadratic in curvature. Results from several independent research groups \cite{Parker69,GribMam70,ZeldPit71,HuFullPar73,FullParHu74,FullPar74, Cr1, Cr3, Alstar1} demonstrate that such terms are linked to the conformal anomaly responsible for particle creation. They found that particle production is closely related to the Weyl tensor and trace anomaly. Consequently, homogeneous and isotropic cosmological models with a vanishing Weyl tensor do not produce particles.

However, the problem of cosmological particle creation has remained unresolved because the back-reaction of created particles and their creation process on the spacetime metric has not been fully explained. The back-reaction problem is particularly complex: solving quantum field equations requires boundary conditions that depend on the spacetime structure, which itself originates from gravitational equations that require an energy-momentum tensor, which in turn depends on quantum field averages.

To circumvent these obstacles, we propose a phenomenological description of particle creation. We employ the action for an ideal fluid in Euler variables \cite{Ray}, where the particle conservation law is replaced by a particle creation law \cite{Ber1}. This method enables us to study particle creation at the classical level while accounting for back-reaction effects.

In our previous works \cite{Uni, Ber6}, we showed that using the conformally invariant action of gravity in combination with our chosen matter action leads to Sakharov's induced gravity \cite{Sakh}. Sakharov was the first to propose that gravity is not a fundamental field existing independently, but rather emerges from the tension of the quantum vacuum of all other quantized fields. This approach is now known as induced gravity, and we adopt this perspective in our work.

Our phenomenological framework enables description of vacuum polarization—it is sufficient to set the creation law to zero, though not identically zero. We will also show that accounting for back-reaction changes the situation drastically. Specifically, the second result of previous investigators is no longer valid: particle creation becomes possible even in homogeneous and isotropic models. Moreover, in our previous articles \cite{Uni,Ber4, Ber6}, we demonstrated that this model suggests particle creation could be responsible for gravitating dark matter and possibly explains dark energy.

In this paper, we focus on studying phase transitions between different vacuum states, as well as transitions from vacuum to non-vacuum states. We will show that, in the most general case, the transition from vacuum to a state where particle production is possible occurs through a light-like hypersurface.

\section{Mathematical tools}
We are working in the framework of the Riemannian geometry and using the units with $\hbar=c=1$. The interval $ds$ between the nearby points of the space-time is determined by the metric tensor $g_{\mu\nu}(x)$
$$ds^2=g_{\mu\nu}dx^\mu dx^\nu$$
with the signature $(+---)$, greek indices run 0, 1, 2, 3. The Riemann curvature tensor is:
$$R^\mu_{\nu\lambda\sigma}=\frac{\partial\Gamma^\mu_{\nu\sigma}}{\partial x^\lambda}-\frac{\partial\Gamma^\mu_{\nu\lambda}}{\partial x^\sigma}+\Gamma^\mu_{\kappa\lambda}\Gamma^\kappa_{\nu\sigma}-\Gamma^\mu_{\kappa\sigma}\Gamma^\kappa_{\nu\lambda},$$
where $\Gamma^\mu_{\nu\lambda}$ are the Christoffel symbols. The Ricci tensor $R_{\mu\nu}$ and the curvature scalar $R$ are as follows:
$$R_{\mu\nu}=R^\sigma_{\mu\sigma\nu}, \quad  R=g^{\mu\nu}R_{\mu\nu}=R^\sigma_\sigma.$$

The Weyl tensor is a completely traceless part of the Riemann curvature tensor,
$$C_{\mu\nu\lambda\sigma}=R_{\mu\nu\lambda\sigma}-\frac{1}{2}R_{\mu\lambda}g_{\nu\sigma}+\frac{1}{2}R_{\mu\sigma}g_{\nu\lambda}-\frac{1}{2}R_{\nu\sigma}g_{\mu\lambda}+\frac{1}{2}R_{\nu\lambda}g_{\mu\sigma}+\frac{1}{6}R(g_{\mu\lambda}g_{\nu\sigma}-g_{\mu\sigma}g_{\nu\lambda}).$$

The notion of local conformal transformation is also important for this paper. It is defined as follows:
$$ds^2=\Omega^2d\hat{s}^2 = \Omega^2(x)\hat{g}_{\mu\nu}dx^\mu dx^\nu,$$
where ``\^{}'' means "transformed", and $\Omega(x)$ is called "the conformal factor". It should be noted that such a transformation does not touch the coordinate system. The Weyl tensor defined above is conformal invariant:
$$C^\mu_{\nu\lambda\sigma}=\hat{C}^\mu_{\nu\lambda\sigma}.$$
We will examine cosmological solutions, specifically the homogeneous and isotropic space-times using the Robertson-Walker metric
$$
ds^2=dt^2-a^2(t)dl^2=a^2(\eta)(d\eta^2-dl^2),
$$
$$dl^2=\frac{dr^2}{1-kr^2}+r^2(d\theta^2+\sin^2\theta d\varphi^2), \quad (k=0,\pm1),
$$
here $t$ is the cosmological time and $\eta$ is the conformal time. Note, that for any scale factor $a$ Weyl tensor is identically zero:
$$C^\mu_{\nu\lambda\sigma}\equiv 0.$$

The Ricci tensor and scalar curvature for the cosmological metric are:
$$
R^0_0=-3\frac{\ddot{a}}{a}, \quad R^1_1=R^2_2=R^3_3,
$$
$$
R=-6\left(\frac{\ddot{a}}{a}+\frac{\dot{a}^2+k}{a^2}\right),
$$
where dot denotes the $t$ derivative.

\section{The model}
The model we propose is based on three postulates:
\begin{enumerate}
    \item The creation of the Universe from "nothing". 
    Alexander V. Vilenkin was the first to propose this idea. By "nothing" we mean vacuum, that is a state without particles which can be "touched" and counted.
    \item Induced gravity. Andrei D. Sakharov proposed in his article \cite{Sakh} that gravity is not a separate fundamental interaction but "simply" the tension of the vacuum of all other quantum fields. Thus, for induced gravity model the total action coincides with the matter action:
    $$
    S_{tot}=S_m .
    $$
    \item Phenomenology.
    The particle creation is a purely quantum process. At the same time we describe gravity with classical equations. To solve a quantum problem, it is necessary to impose boundary conditions, for which in turn we need to know the global structure of space-time. The space-time geometry is determined by classical field equations, where the source is the averaged energy-momentum tensor of these very quantum fields. That is, we need a solution to the quantum problem at the first place. To overcome this obstacle we considered a phenomenological description of the crated particles in the form of ideal fluid. Our model also takes into account the back reaction on the metric not just from the energy-momentum tensor of the produced particles but also from the creation process itself and vacuum polarization. 
\end{enumerate} 

In accordance with the above postulates, we will consider the action of matter, which is a modification of the action of an ideal fluid in Eulerian variables \cite{Ray} first presented in the article \cite{Ber1}: 
\begin{eqnarray}
S_{\rm m}&=& -\!\int\!\varepsilon(X,n,\varphi)\sqrt{-g}\,d^4x + \int\!\lambda_0(u_\mu u^\mu-1)\sqrt{-g}\,d^4x+\nonumber\\
&&+\int\!\lambda_1\left((n u^\mu)_{;\mu}-\Phi\right)\sqrt{-g}\,d^4x+ \int\!\lambda_2 X_{,\mu}u^\mu\sqrt{-g}\,d^4x.
\end{eqnarray}
Here the semicolon ";" denotes the covariant derivative. The dynamical variables are the particle number density $n(x)$, the vector field $u^\mu(x)$, scalar field $\varphi$ and the auxiliary variable $X(x)$ , while $\lambda_0(x)$, $\lambda_1(x)$ and $\lambda_2(x)$ are Lagrange multipliers. Lagrange multipliers provide us with three constraints:
$$u^\sigma u_\sigma - 1 = 0$$
makes the vector field $u^\mu$ similar to the four-velocity of particles,
$$(nu^\sigma)_{;\sigma}=\Phi[inv]$$
manifests the so called particle creation law, and
$$X_{,\mu}u^\mu=0$$
links the vector field $u^\mu$ to the particle trajectories and enumerates them.

In the homogeneous and isotropic cosmological models the Robertson-Walker frame of reference is supposed to be comoving, therefore $u^t=u_t=1, u^i=0$ $(i=1,2,3)$ for the cosmological time $t$, and $u^\eta=\frac{1}{a}$, $u_\eta=a$, $u^i=0$ for the the conformal time $\eta$. Note, that if $a=0$ for some $t=t_0$ or $\eta=\eta_0$ the "initial" time surface $t=t_0$ ( $\eta=\eta_0$) is also null not the spacelike as it is used to think of.

The creation function $\Phi[\rm inv]$ consists of some invariants dependent on geometry and some fields whose quanta are supposed to be created. Fortunately, it is possible to be more precise. Let us show that the left hand side of the creation law is conformal invariant:
$$
(nu^\mu)_{;\mu} = \frac{(nu^\mu\sqrt{-g})_{,\mu}}{\sqrt{-g}}=\frac{\left(\frac{\hat{n}}{\Omega^3}\frac{\hat{u}^\mu}{\Omega}\Omega^4\sqrt{-\hat{g}}\right)_{,\mu}}{\sqrt{-g}} = \frac{(\hat{n}\hat{u}^\mu\sqrt{-\hat{g}})_{,\mu}}{\sqrt{-g}},
$$
where we used the rules for conformal transformation for the corresponding variables:
$$n=\frac{\hat n}{\Omega^3},  \quad u^\mu=\frac{\hat u^\mu}{\Omega},  \quad \sqrt{-g}=\Omega^4 \sqrt{-\hat g} \,.$$
Hence, $\Phi\sqrt{-g}$ must be conformal invariant. Here we will consider only the particle production by some scalar field $\varphi$. Note that $\varphi$ is not a completely ordinary classical scalar field, it has the "wrong" sign of the kinetic term, which allows particles to be produced. It rather resembles the C-field first introduced by F. Hoyle \cite{Hoyle, Hoyle1}. 

Taking into account the conformal invariance shown above, the source function has the following form:
\begin{equation}
\Phi = \alpha C^2 + \beta\left(\varphi\Box\varphi - \frac{1}{6}\varphi^2 R + \Lambda\varphi^4\right) + \varepsilon_1(\varphi, n),
\end{equation}
where $\alpha,\beta,\Lambda$ are some constants, $\square$ stands for d’Alambertian. Let us explain the meaning of each term here:
\begin{itemize}
\item The square of the Weyl tensor $C^2$ represents a pure geometrical contribution. 
\item We suggest that the particles are created by quantum fluctuations of some scalar field $\varphi$. Assuming its usual transformation under the local conformal transformation $\varphi =\frac{\widehat{\varphi }}{\Omega }$, we get the second term in $\Phi$.
\item Our particles are just mass shell quanta of the scalar field. They may annihilate back to the virtual particles with the rate depending on the particle number density. Therefore, there should be corresponding contribution to the creation law: $\varepsilon_1(\varphi, n)$.
\end{itemize}

It is important to emphasize that in our model the particles can be created in homogeneous and isotropic manner. Its gravitational influence may appear much more stronger than that of ordinary (visible) matter created by the inhomogeneous and anisotropic fluctuations considered for example by V. F. Mukhanov and G. V. Chibisov \cite{Muhanov Chibisov}. 

Next we would like to investigate the structure of the energy density $\varepsilon$ and the term $\varepsilon_1$ in the creation law. To do this, let us consider their behavior under the local conformal transformation. Due to the adopted induced gravity point of view, $S_{tot}=S_m$. Surely, the action integral does not need to be conformal invariant, but its variation does because this transformation does not touch the coordinate system (and hence, the limits of the integration). Indeed,
$$\frac{\delta S}{\delta \Omega}=\frac{\delta S}{\delta \psi}\frac{\delta \psi}{\delta \Omega}\equiv0,$$
($\psi$ is the collective dynamical variable). It is easy to prove that the terms with the Lagrange multipliers in our action $S_m$ are conformal invariant and the Lagrange multipliers themselves are assumed conformal invariant. So, we have to check only the first term,
$$\frac{\delta}{\delta\Omega}\int \varepsilon (X,\varphi, n) \sqrt{-g}\,  d^{4}x.$$
Since the enumeration of the trajectories, $X$, is not touched by the conformal transformation, and $n=\frac{\hat n}{\Omega^3},$ $u^\mu=\frac{\hat u^\mu}{\Omega},$ $\sqrt{-g}=\Omega^4 \sqrt{-\hat g}$, we get:
\begin{equation} \label{eq1}
\varphi\,  \frac{\partial \varepsilon }{\partial \varphi }+3n\, \frac{\partial \varepsilon }{\partial n}=4\, \varepsilon.   
\end{equation}
The general solution of this equation has the form 
$$\varepsilon =F\left ( x \right )\, \varphi ^{4}, \quad x = \frac{n}{\varphi ^{3}}$$
with $F(0)=\sigma$, therefore $\sigma \, \varphi ^{4}$ can be considered as the vacuum energy density. 

Two important examples are the dust-like matter:
$$
\varepsilon=\mu_1 \,n \varphi,
$$
and the thermal bath:
$$
\varepsilon=\mu_2\, n^{\frac{4}{3}},
$$
where $\mu_1$ and $\mu_2$ are some constants. Note, that the mass of the dust particles depends on $\varphi$. 

Surely, the structure of the $\varepsilon_1$-term is the same
$$
\varepsilon_1 =F_1\left ( x \right )\, \varphi ^{4},
$$
with $F_1(0) = 0$, since the $\varphi^4$-term is already present in the conservation law.

We'll call the term $\varepsilon_1$ in the creation law "gravitating mirages" \cite{Uni}. It contributes to the energy-momentum tensor through a Lagrange multiplier $\lambda_1$. While $F(x)$ describes actual particles (like dust proportional to $x$), $F_1(x)$ can have different behavior (like $x^{\frac{4}{3}}$ for thermal baths) and represents something that gravitates but isn't matter. These mirages represent the gravitational back-reaction of particle creation on spacetime geometry, forming an invisible component of the gravitating system. Let us also emphasize that $F_1(x)$ is given in the Lagrangian, but $F(x)$ appears as a result of solving the equations of motion, because there is the creation of particles, but it is not known in advance what type exactly.
 
For convenience, we will divide the action into two parts:
$$S_m=S_m[n]+S_m[\varphi]$$
\begin{eqnarray}
    S_m[n] &=& -\int F(x)\varphi^4\sqrt{-g}d^4x+\int\lambda_0(u^\sigma u_\sigma-1)\sqrt{-g}d^4x+\\
    &&+\int\lambda_1((nu^\sigma)_{;\sigma}-F_1(x)\varphi^4)\sqrt{-g}d^4x+\int\lambda_2X_{,\sigma}u^\sigma\sqrt{-g}d^4x, \nonumber
\end{eqnarray}
$$S_m[\varphi]=-\int \lambda_1(\alpha C^2+\beta(\varphi\varphi^\sigma_{;\sigma}-\frac{1}{6}\varphi^2R+\Lambda\varphi^4))\sqrt{-g}d^4x.$$

We ignore here the dependence of function F on the auxiliary dynamical variable $X$, because its variation gives us the equation for the Lagrange multiplier $\lambda_2$ which of no use in cosmology. For the same reason we will put $C^2=0$ in the action integral because $C\delta C=0$.
Let's get the equations of motion for our action. The variation of the particle number density, $\delta n$, gives us the following equation:
$$
-\frac{dF}{dx}\varphi-\lambda_{1,\sigma}u^\sigma-\lambda_1\frac{dF_1}{dx}\varphi = 0,
$$
surely, we integrated by parts the term $\lambda_1(nu^\sigma)_{;\sigma}\sqrt{-g} = \lambda_1(nu^\sigma\sqrt{-g})_{,\sigma}$ and put to zero the corresponding surface term. By varying the action on the dynamic variable $\delta u^\mu$ we obtain,
$$
2\lambda_0u_\mu-\lambda_{1,\mu}n+\lambda_2X_{,\mu} = 0,
$$
and for $\delta X$ we have:
$$\frac{\partial \varepsilon}{\partial X}-(\lambda_2u^\sigma)_{;\sigma} = 0.$$
By contracting the second (vector) equation with $u^{\mu}$, using constraints and the first (scalar) equation, we calculate the Lagrange multiplier $\lambda_0$:
$$
2\lambda_0 = -x\left( \frac{dF}{dx}+\lambda_1\frac{dF_1}{dx} \right)\varphi^4.
$$
The energy-momentum tensor is also divided into two parts:
$$
T^{\mu\nu}[n] = \varepsilon g^{\mu\nu}-2\lambda_0 u^\mu u^\nu + g^{\mu\nu}(n\, \lambda_{1,\sigma}u^\sigma+\lambda_1F_1(x)\varphi^4),
$$
substituting for $\lambda_0$ the expression found above gives us
\begin{eqnarray}
    T^{\mu\nu}[n] &=& x\varphi^4\left(\frac{dF}{dx}+\lambda_1\frac{dF_1}{dx}\right)u^\mu u^\nu-\\
    &&- \varphi^4\left(x\frac{dF}{dx}-F+\lambda_1\left(x\frac{dF_1}{dx}-F_1\right)\right)g^{\mu\nu},
\end{eqnarray}
and the remaining part of the energy-momentum tensor is
\begin{eqnarray}
    T^{\mu\nu}[\varphi] = \beta\left( (\lambda_1\varphi)^{;\mu}\varphi^\nu + (\lambda_1\varphi)^{;\nu}\varphi^\mu - ((\lambda_1\varphi)_{,\sigma}\varphi^\sigma - \lambda_1\Lambda\varphi^4)g^{\mu\nu} \right)\\
    + \frac{\beta}{3}\left( \lambda_1\varphi^2\left(R^{\mu\nu} - \frac{1}{2}Rg^{\mu\nu}\right) -(\lambda_1\varphi^2)^{;\mu;\nu} + (\lambda_1\varphi^2)^{;\sigma}_{;\sigma}g^{\mu\nu}\right).\nonumber
\end{eqnarray}

The variation of the scalar field, $\delta\varphi$, gives the following equation:
\begin{eqnarray}
    \left(3x\frac{dF}{dx}-4F\right)\varphi^3+\lambda_1\left(3x\frac{dF_1}{dx}-4F_1\right)\varphi^3 =\\
    = \beta\left(\lambda_1\varphi^\sigma_{;\sigma} + (\lambda_1\varphi)^{;\sigma}_{;\sigma}+4\lambda_1\Lambda\varphi^3-\frac{1}{3}\lambda_1\varphi R\right). \nonumber
\end{eqnarray}
Because of the induced gravity,
$$
T^{\mu\nu}[n] + T^{\mu\nu}[\varphi] = 0 \quad \Rightarrow \quad T=g_{\mu \nu} T^{\mu \nu}=0
$$
This condition is equivalent to the equation (\ref{eq1}) obtained above:
\begin{multline} \nonumber
T=\varepsilon-3p+4\beta \,  \lambda _{1}\, \Lambda\, \varphi ^{4}-\frac{\beta }{3}\, \lambda _{1}\, \varphi ^{2}\, R+\beta \,  \varphi \, \square \left ( \lambda _{1}\varphi  \right )+\beta  \, \lambda _{1}\varphi\,  \square \varphi+\\+\lambda _{1}\varphi^{4}\left ( 4F_1-3x\frac{\mathrm{d} F_1}{\mathrm{d} x} \right )=\varepsilon -3p-\varphi \, \frac{\partial \varepsilon}{\partial \varphi }=0, 
\end{multline}
where in the second equality the equation of motion obtained by variation in $\varphi$ was used. Since $T^{1}_{1}=T^{2}_{2}=T^{3}_{3}$ for cosmological metric, we can use only $T^{00}$ and $T$ but as we have just shown, the condition $T=0$ gives us an equation that determines the general form of the function $\epsilon$, so all that remains is the equation $T^{00}=0$. 
EOM can be expressed in terms of conformally invariant functions $N=n a^3$, $\lambda_1(\eta)$, $f=\varphi a$ and conformal time $\eta$:
\begin{eqnarray} \label{1}
\lambda'_{1}\,ff'+\lambda_{1}\left (f'^2+kf^2+\Lambda f^4  \right )+\frac{f^4}{\beta }\left (F+\lambda _{1}\,F_1  \right )=0,
 \\ \nonumber
 f\lambda _{1}''+2\lambda_1'f'+2\lambda _{1}\left ( f''+kf+2\Lambda f^{3} \right )+\label{2}\\+\frac{1}{\beta }\left ( 4f^3F-3N \frac{dF}{dx}+\lambda _{1}\left (4f^3F_1-3N \frac{dF_1}{dx}   \right ) \right )=0, 
\\ 
\beta f\left ( f''+kf+\Lambda f^3 \right )+f^4 F_1=N', \label{3}  \\ \label{4}
\left (\frac{dF}{dx}+\lambda _{1} \frac{dF_1}{dx}  \right ) f +\lambda '_{1}=0, \end{eqnarray}
where prime denotes the derivative with respect to $\eta$, $F,F_1$ are considered as functions of the combination $x=\frac{N}{f^3}$. It should be noted here that our equations in conformal variables are valid only for $a \neq 0$.

The second equation can be reproduced by differentiating of the fourth equation and making use of the first and second ones, what is the reflection of the conservativity of the Einstein tensor. Thus, we have 3 equations for 3 conformal invariant functions, $\lambda_1(\eta)$, $x(\eta)$ and $f(\eta)$.
The scale factor $a(\eta)$ does not enter these equation as it should be, because it plays, actually, the role of the conformal factor $\Omega$. Therefore, the only trace of the geometry is the type of the cosmological model encoded in the value of $k$, namely, $k=1$ for the closed universe, while $k=0$ for the spatially flat and $k=-1$ for the open ones. In order to obtain the specific $a(\eta)$, we should impose the gauge fixing condition.
\section{Phase transition}
When deriving the equations of motion, the dynamic variable $n$ and therefore $N$ were considered continuous, but from a physical point of view it is discrete, since $N=na^3$ is a number of particles in a unit of conformal volume. Thus, the dependence of N on time should look like this:
\begin{equation*}
N=N_0\, \theta(\eta-\eta_0)+(N_1-N_0)\,\theta(\eta-\eta_1)
+(N_2-N_1)\,\theta(\eta-\eta_2)+...,    
\end{equation*}
that is, the number of particles changes in jumps and remains constant between moments $\eta_i$ in time. We will call such time intervals between $\eta_i$ and $\eta_{i+1}$ when the number of particles is constant intermissions.

Let us consider one such transition at some point in time $\eta_0$:
\[
N = N_+ \theta(\eta-\eta_0) + N_- \theta(\eta_0-\eta)
\]
(+) = after, (-) = before,
$\eta =\eta_0$ is an act of creation, $N_{\pm}=const$ during the intermissions. 
\begin{equation*}
N' = [N]\delta(\eta-\eta_0), 
\end{equation*}
The functions $f(\eta)$, $a(\eta)$, $ \lambda _{1}(\eta)$ should be continuous on the hypersurface $\eta=\eta_0$:
$$
[f]=0, \quad  [\lambda_1]=0
$$
in order to avoid the occurrence of undefined functions, but their first derivatives in general may contain jumps, and the second may contain a delta function.
Comparison of $\delta$-functions:
\begin{align}
[N] = \beta f [f'], \label{N}\\
2\lambda_1 [f'] + f \left[ \lambda_1' \right]=0.
\end{align}
Let $f \neq 0$ on the boundary ($\eta=\eta_0$) $\Rightarrow$
\begin{align}
2\lambda_1 \beta f \left[f' \right] + \beta f^2 \left[ \lambda_1' \right] = 0 \Rightarrow 
2\lambda_1 [N] + \beta f^2 \left[ \lambda_1' \right] = 0
\end{align}
Our trick: we fix everything "before" on the boundary then from the last equation we have:
\begin{equation}
2\lambda_1 N + \beta f^2  \lambda_1' =2\lambda_1 N_{-} + \beta f^2  \lambda_{1-}'.  
\end{equation}
Note that both $\lambda_1$ and $f$ are also fixed. Then using equation (\ref{4}) we obtain:
\begin{align}
2\lambda_1 x - \beta \left\{ \lambda_1 \frac{dF_1}{dx} + f \frac{dF}{dx} \right\} = 2\lambda_1 x - \beta \left\{ \lambda_1 \frac{dF_1}{dx}(-) + f\frac{dF}{dx}(-) \right\} \Rightarrow\\ \label{5}
\Rightarrow \beta \left\{ \lambda_1 \frac{dF_1}{dx} + f \frac{dF}{dx} \right\} = 2\lambda_1 x + C_-,
\end{align}
where we introduced the notation:
\begin{equation*}
C_- = \beta \left\{ \lambda_1 \frac{dF_1}{dx}(-) + f \frac{dF}{dx}(-) \right\} - 2\lambda_1 x_- .
\end{equation*}
Next, we integrate equation (\ref{5}) with respect to $x$ while remaining on the boundary $\eta=\eta_0$. That is, we change the number of particles $N(x)$ after the transition, considering it as a parameter, but the number of particles before the transition $N_{-}$ remains fixed:
\begin{equation} \label{6}
\beta(\lambda_1 F_1 + F) = \lambda_1 x^2 + C_- x + C_+,
\end{equation}
here $C_+ = \beta\sigma$ on the boundary. Note, that function $F$ enters all equations in exactly this combination.

Let's use this result for the equation (\ref{2}) (on the boundary)
\begin{align}
 f^3 \left\{ 2\lambda_1 x^2 - C_- x - 4\beta\sigma \right\} = \beta^2 \left\{ 2\lambda_1 f'' + f \lambda_1'' + 2 \lambda_1' f' + 2\lambda_1 kf + 4\Lambda f^3 \right\} .
\end{align}
From the equation (\ref{3}) it follows (for $f \neq 0$):
\begin{equation}
f''=-kf^2 - \Lambda f^3 - \frac{1}{\beta} F_1(x) f^3,
\end{equation}
substituting it into the previous equation, we get:
\begin{equation} \label{7}
2\lambda_1 x^2 - C_- x - 4\beta\sigma - 2\beta^2 \Lambda \lambda_1 = \frac{\beta^2}{f^3} ( f  \lambda_1'' + 2 \lambda_1' f') - 2\beta \lambda_1 F_1  . 
\end{equation}
Let's differentiate equation (\ref{4}) with respect to $\eta$ for $N=const$:
\begin{align}
\lambda_1''= f'\left\{ 3x \left( \lambda_1 \frac{d^2 F_1}{dx^2} + f \frac{d^2 F}{dx^2} \right) - \left( \lambda_1 \frac{dF_1}{dx} + f \frac{dF}{dx} \right) \right\} + f^2 \frac{dF_1}{dx} \left\{ \lambda_1 \frac{dF_1}{dx} + f \frac{dF}{dx} \right\},  \end{align}
substituting (\ref{6}) in this equation we obtain
\begin{equation*} \label{8}
\beta \lambda_1'' = f'\beta \left\{ 4\lambda_1 x - C_- \right\} + f^2 \frac{dF_1}{dx} \left\{ 2\lambda_1 x + C_- \right\} .
\end{equation*}
Let's use it in the equation (\ref{7}):
\begin{align} \label{9}
2\lambda_1 x^2 - C_- x - 4\beta\sigma - 2\beta^2 \Lambda + 3\beta \frac{C_-}{f^2} f' = \beta \left\{ \frac{dF_1}{dx}(2\lambda_1 x + C_-) - 2\lambda_1 F_1 \right\} .
\end{align}
Let's repeat our trick once more to find $\frac{df}{dy}$:
\begin{align}
N-N_- = \beta f f' - \beta f f'_{-} \Rightarrow\\ \label{f}
\Rightarrow \frac{\beta}{f^2} f' = x+\alpha,
\end{align}
here we introduced the notation:
\begin{equation} \label{al}
\alpha \equiv \frac{\beta}{f^2}f'_{-} - x_{-}\,  .   
\end{equation}
Substituting $f'$ into equation (\ref{9}) we obtain a first-order linear differential equation for $F_1(x)$
\begin{equation} \label{F1}
\beta \left\{ \left(x+\frac{C_-}{2\lambda_1}\right) \frac{dF_1}{dx} - F_1 \right\} = x^2 + \frac{C_-}{\lambda_1}x + \frac{3C_-\alpha}{2\lambda_1} - \frac{2\beta\sigma}{\lambda_1} - \beta^2\Lambda .  \end{equation}

We use relation (\ref{6}) to express $F+\lambda_1 F_1$, $\lambda_1'$ from (\ref{4}), the expression for $f'$ from (\ref{f}), and substitute it all into equation (\ref{1}) at the boundary:
\begin{equation} \label{alC}
\alpha^2 - \frac{C_-}{\lambda_1}\alpha + \frac{\beta\sigma}{\lambda_1} + \beta^2 \left( \frac{k}{f^2} + \Lambda \right) = 0,
\end{equation}
that is, we obtained a ratio linking $\alpha$ and $C_{-}$.

To move on, let's find the solutions for $F_1(x)$ from (\ref{F1}). This equation can be rewritten as follows:
\begin{align}
\left(x+\frac{C_-}{2\lambda_1}\right) \frac{d}{dx}(\beta F_1) - (\beta F_1) = \left(x+\frac{C_-}{2\lambda_1}\right)^2 + A,\\
A = \frac{3}{2\lambda_1} C_- \alpha - \frac{2\beta\sigma}{\lambda_1} - \beta^2\Lambda - \frac{C_-^2}{4\lambda_1^2},
\end{align}
then the solution for $F_1(x)$ is:
\begin{equation*}
\beta F_1 = x\left(x + 3\alpha - \frac{4\beta\sigma}{C_-} - \frac{2\beta^2\Lambda}{C_-} \lambda_1 \right), \quad C_-\neq 0,
\end{equation*}
\begin{equation*}
\beta F_1 = x(x+\zeta_0), \quad C_-=0 \Rightarrow A=0 \Rightarrow 2\sigma = -\beta\Lambda \lambda_1,
\end{equation*}
where $\zeta_0$ is an undefined constant.

Now with $F_1(x)$ known let's find $F(x)$ from (\ref{6}):
\begin{align} \label{F}
\beta F = x\left\{ C_- - 3\alpha\lambda_1 + \frac{4\beta\sigma\lambda_1}{C_-} + \frac{2\beta^2\Lambda\lambda_1^2}{C_-} \right\} + \beta\sigma, \quad C_-\neq 0, \\
\beta F = -\lambda_1 \zeta_0 x + \beta\sigma, \quad C_-=0.
\end{align}
Let's summarize our results so far:
\begin{align}
    f \neq 0,\\
    \beta \lambda_1' = -f(2\lambda_1 x + C_-),
\end{align}
\begin{align}
\left\{
\begin{aligned} \label{FsC}
\beta F_1 &= x^2 + x\left(3\alpha - \frac{4\beta\sigma}{C_-} - \frac{2\beta^2\Lambda}{C_-} \lambda_1\right) ,\\
\beta F &= x\left\{ C_- - 3\alpha\lambda_1 + \frac{4\beta\sigma}{C_-}\lambda_1 + \frac{2\beta^2\Lambda}{C_-}\lambda_1^2 \right\} + \beta\sigma, \quad C_{-} \neq 0,
\end{aligned}
\right.
\end{align}

\begin{align} \label{Fs}
\left\{
\begin{aligned}
\beta F_1 &= x(x+\zeta_0), \\
\beta F &= -\lambda_1 \zeta_0 x + \beta\sigma, \quad C_{-}=0.
\end{aligned}
\right.    
\end{align}

Thus, we have obtained a fixed form for $F_1$ and $F$ as functions of $x$. In particular, this means that within the framework of our model, only dust particles can be produced.

\section{Vacua}
\begin{figure}
 \centering
        \includegraphics[width=0.7\textwidth]{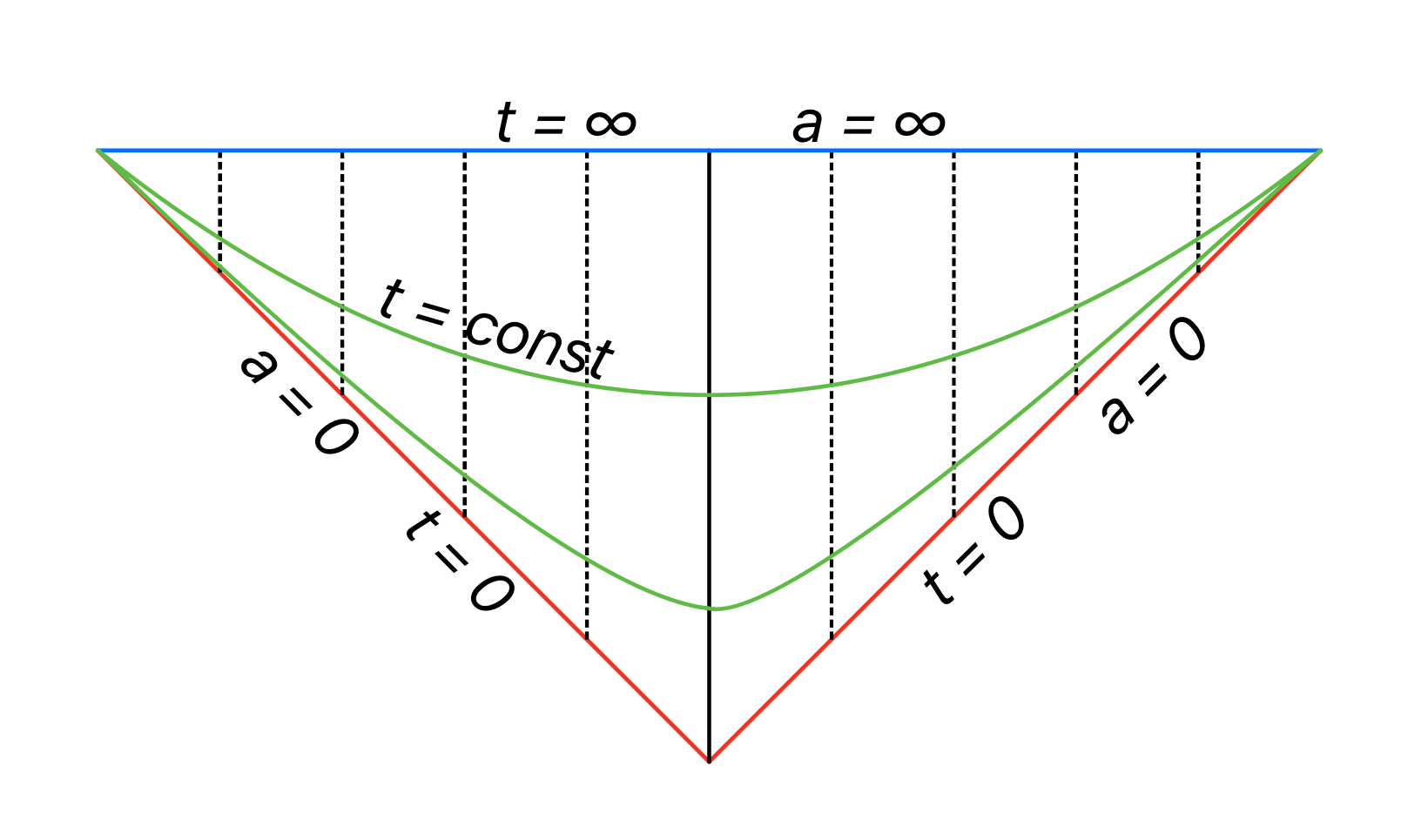}
 \caption{Diagram of the cosmological model.}
\label{fig1}
\end{figure}

Particle creation implies that system exits from the vacuum. By vacuum we understand the solution with no particles and no particle creation. Thus,
$$
N=0, \quad N'=0 .
$$
Let us introduce the constants: $F(0)=\sigma,\quad \frac{dF}{dx}(0)=\mu_1,\quad  \frac{dF_1}{dx}(0)=\gamma_1$, and we remember that $F_1(0)=0$. For the vacuum case the EOM become:
$$
\lambda_1'+f \left(\mu_1+\lambda_1 \, \gamma_1 \right)=0,
$$
$$
 f\left\{ f''+k \, f+\Lambda \, f^3\right\}=0,
$$
$$
\sigma f^{4}+\beta \left\{  ( \lambda_1 f )'f'+\lambda_1 \, k f^2+\lambda_1\, \Lambda f^4\right\}=0,
$$
here we did not include in this set the equation obtained by variation $\delta \varphi$, since it is the differential consequence of all other equations. Note, that we have 3 equations for only 2 functions, $\lambda_1(\eta)$ and $f(\eta)$, so the set of the equations is overdetermined. The second equation is splittеd naturally into two branches: either $f \equiv 0$, or
$$
f''+k \, f+\Lambda \, f^3=0.
$$
We divide vacua into classical and quantum. For classical vacua we can extend the classical equations unambiguously through the space-like hypersurface $t=0$ ($\eta=0$). For the
quantum vacuum we cannot uniquely extend EOM and the conformal coordinate system, because this hypersurface is light-like (see Fig.\ref{fig1}). In the quantum vacuum region it is possible to introduce formally, in the absence of any observers, the analog of the Robertson-Walker metric with conformal time, but one can not be sure about the overall conformal factor and about the signature of the whole manifold (assumingly 4-dimensional) which can be $(+ - - -)$ as well as $(- - - -)$ and even $(+ + - -)$ (see \cite{Ash} as an example of using mixed signature in cosmology). In our case: the quantum vacuum is $a=0, t=0$ and it corresponds to $f=0$. 

There are as many as 3 classical ones:
\begin{itemize}
\item {Vacuum $\#1$: \quad $f(\eta) \equiv 0, \quad a\neq 0, \quad \varphi=0,\quad  \lambda_1 = const;$}
\item{ Vacuum $\#2$: 
\begin{eqnarray} \nonumber
f=f_0=\pm \sqrt{-\frac{k}{\Lambda}}, \quad \lambda_1(\eta )=\left ( \lambda_1(0)+\frac{\mu_1}{\gamma_1} \right )e^{-f_0\gamma_1 \eta }-\frac{\mu_1}{\gamma_1} ,\quad \sigma=0;    
\end{eqnarray}
}
\item{Vacuum $\#3$: \quad
$
\lambda_1=-\frac{\mu_1}{\gamma_1}, \quad  \sigma=\frac{1}{2}\frac{\beta \mu_1}{\gamma_1}\Lambda. 
$}
\end{itemize}

Vacuum $\#3$ does not require the special value for $\lambda_1$. In contrast, the value of $\lambda_1$ determines the characteristics of the particles created. The special value of $\sigma$ means the choice of the solution (since $\sigma$ is just the integration constant). For vacuum $\#3$ , the function $f$ satisfies the equation:
\begin{equation*}
f'^2+U(f)=0, \quad U(f)=kf^2+\frac{1}{2}\Lambda f^4,
\end{equation*}
with the solutions:
\begin{itemize}
\item{ $k = 0,\, \Lambda < 0$: \quad  $f^2 = -\frac{2}{\Lambda}\frac{1}{(\eta-\eta_0)^2};$ }
\item {$k = 0, \, \Lambda = 0$: \quad $f=const;$ }
\item {$k = -1,\,  \Lambda > 0$: \quad $f^2 = \frac{2}{\Lambda}\frac{1}{ch^2(\eta-\eta_0)};$}
\item{$k = -1,\, \Lambda < 0$: \quad $f^2 = -\frac{2}{\Lambda}\frac{1}{sh^2(\eta-\eta_0)};$}
\item{$k = -1,\, \Lambda = 0$: \quad $f= e^{\pm(\eta-\eta_0)};$}
\item{$k = 1,\, \Lambda < 0$: \quad $f^2 = -\frac{2}{\Lambda}\frac{1}{sin^2(\eta-\eta_0)}.$}
\end{itemize}

Note that if $\gamma_1=0$, then vacuum $\#3$ exists only under the condition that $\mu_1=0$. In that case:
\begin{equation*}
 \lambda_1=const,  \quad \sigma=-\frac{1}{2}\beta \Lambda \,\lambda_1.    
\end{equation*}

The Table \ref{tab1} shows whether a particular vacuum exists depending on the value of the parameters. Here "$+$" means that the vacuum exists, "$-$" means it does not exists, "$\equiv$" means it coincides with another vacuum.

\begin{table}[h]
\caption{Dependence of the existence of vacua on the parameters $\Lambda$ and $k$.}
\centering
\begin{tabular}{l c c c c}
\hline
$k$ & $\Lambda$ & $\#1$ & $\#2$ & $\#3$ \\
\hline
$0$ & $>0$ & $+$ & $\equiv \#1$ & $-$ \\
$0$ & $<0$ & $+$ & $\equiv \#1$ & $+$ \\
$0$ & $=0$ & $+$ & $+$ & $+$ \\
$1$ & $>0$ & $+$ & $-$ & $-$ \\
$1$ & $<0$ & $+$ & $+$ & $+$ \\
$1$ & $=0$ & $+$ & $\equiv \#1$ & $-$ \\
$-1$ & $>0$ & $+$ & $+$ & $+$ \\
$-1$ & $<0$ & $+$ & $-$ & $+$ \\
$-1$ & $=0$ & $+$ & $\equiv \#1$ & $+$ \\
\hline
\end{tabular}
\label{tab1}
\end{table}

Although the parameter $k$ can be either $0$ or $\pm 1$ in our model, physical considerations dictate that any classical vacuum must correspond to $k = +1$, since the probability of an infinite fluctuation in space is infinitely small. Therefore, when further considering classical vacua, we will assume that $k = 1$.

Since for vacuum $\#1$ $\varphi=0$, there is no one to give birth therefore there is no transition to particle creation. This is also evident from the equation (\ref{N}), since with $f=0$ and $[N]\neq 0$ we get $\frac{df}{d \eta} = \infty$ in all space simultaneously, which is improbable. However there is tunneling into vacuum $\#3$. There cannot be classical transitions between different vacua because the functions $\lambda_1(\eta) ,\; f(\eta),\; a(\eta)$ must be continuous under the transition. Nevertheless there is one case of the quantum tunneling between two of them. For $k=1, \; \Lambda<0$ vacuum $\#1$ may undergo the quantum tunneling into the vacuum $\#3$ with $\lambda_1=const$, thus determining the value of $\mu_1$. Such a process begins at the past infinity of the imaginary time $i \eta$. Since the period of the imaginary time is infinite, the temperature after this transition is zero.

From the definitions of $C_-$ and $\alpha$, it is clear that in vacuum \#2 $\alpha=0$, and in vacuum \#3 $C_{-}=0$.

\section{First intermission}
Let's consider the first intermission after exiting the vacuum. Formulas (\ref{FsC}, \ref{Fs}) obtained earlier provide a dependence $F$ and $F_1$ on $x$ even outside the transition surface; in general, they can be presented as follows:
\begin{align}
F=\mu_1 \, x+\sigma,\\
F_1=\frac{1}{\beta}x^2+\gamma_1 \, x,
\end{align}
using these relations, as well as the fact that $N=N_0=const$ in the intermission we obtain from the equation of motion (\ref{3})
\begin{align} \label{df}
f'^2=-kf^2-\frac{1}{2}\Lambda f^4+\frac{N_0^2}{\beta^2 f^2}-\frac{2N_0 \gamma_1}{\beta} f+C_0, \end{align}
where $C_0$ is some constant of integration. 

As noted above, equation of motion (\ref{2}) can be obtained from the others, but even if we consider only the three remaining equations of motion, the system turns out to be overdetermined, since we have three equations for two unknown functions $f(\eta)$ and $\lambda_1(\eta)$. 

If $f=const$, then from equations (\ref{1}) and (\ref{3}) we have:
\begin{align}
\beta \left(k f^2+\Lambda f^4\right)+F_1(x) f^4=0,  \\
f^4 F=0,
\end{align}
since $F\geq 0$ because $\varepsilon=F \varphi^4$ and we do not consider exotic matter then either $f=0$ or $F=0 \Rightarrow\sigma=N_0=0$. Thus, in this case, it is not possible to exit from any of the classical vacua. 

Let us now consider the case when $f \neq const$, then we can move from the variable $\eta$ to $f$. In this case, from (\ref{1}) and (\ref{4}) we obtain two different first-order differential equations for the same function $\lambda_1$:
\begin{align} \label{eq}
\frac{d}{df}\lambda_1+ a_1 \lambda_1+b_1=0,\\
\frac{d}{df}\lambda_1+ a_2 \lambda_1+b_2=0,
\end{align}
where
\begin{align}
a_1=\frac{f}{f'}\frac{d}{dx}F_1,\quad
a_2=\frac{1}{\beta f f'^2}(f^4 F_1+\beta f'^2),\\
b_1=\frac{f}{f'}\frac{d}{dx}F,\quad
b_2=\frac{f^3}{\beta f'^2}F,
\end{align}
from here it follows that either 
$$
a_1=a_2, \quad b_1=b_2,
$$
then, taking into account (\ref{df}), we obtain:
\begin{align}
C_0=0, \quad -\gamma_1=\frac{\sigma}{\beta \mu_1}, \quad -\frac{1}{2}\Lambda=\frac{\sigma^2}{\beta^2 \mu_1^2}, \quad k=0,\\
f'=\frac{1}{\beta}\left(\frac{N_0}{f}+\frac{\sigma}{\mu_1}f^2\right).
\end{align}
that is, the exit is possible only from vacuum $\#3$ since $\sigma \neq 0$ but only for a certain selection of parameters in the Lagrangian: $\Lambda=-2 \gamma_1, \quad k=0$. However, as it was mentioned before, classical vacuum corresponds to $k=1$.

The second possibility is that the following condition is satisfied: 
\begin{align}
\frac{a_2b_1-a_1 b_2}{a_1-a_2}=\frac{d}{df} \left(\frac{b_2-b_1}{a_1-a_2}\right),
\end{align}
it follows from equations (\ref{eq}) if we independently express $\lambda_1$ and $\frac{d}{df}\lambda_1$ through the coefficients $a_i$ and $b_i$ $(i=1,2)$. We will not present this condition in expanded form, as it turns out to be too cumbersome, but we have verified that it is satisfied for any $f$ only if the coefficients $a_i$ and $b_i$ coincide. Otherwise, it yields an additional algebraic equation for the function $f$, from which we obtain that $f=const$ and we've already discussed this case before.

Summarizing all of the above, we find that exiting classical vacua within the framework of the model under consideration is impossible.

\section{Discussion and Conclusion}
In this work, we investigate cosmological particle production arising from vacuum fluctuations of a scalar field, where "cosmological" refers to homogeneous and isotropic processes. We describe the particle creation phenomenologically to account for back-reaction effects, ensuring self-consistency in our approach.

We successfully reformulated the cosmological field equations using only conformally invariant dynamical variables. Because the scale factor in homogeneous and isotropic cosmological models acts as the conformal factor, it vanishes from the resulting system of equations.

A significant feature of our model is the emergence of "gravitating mirages"—terms involving $F_1$ that appear in the energy-momentum tensor. While not conventional matter, these terms produce gravitational effects and may provide an explanation for dark matter, particularly since they can possess negative "energy density" when $\lambda_1<0$. These features can therefore be employed in interpreting observational data, specifically through the "GR-gauge" framework \cite{Ber6}, which reproduces General Relativity in cosmological contexts.

By exploiting the homogeneity and isotropy of the cosmological metric, the conformal invariance of the creation law, and the discrete nature of the particle number $N$, we determined the general functional forms governing the equation of state for both real produced particles and "mirage particles": $F(x)=\sigma+\mu_1 x$ and $F_1=\frac{1}{\beta}x^2+\gamma_1 x$.

We demonstrate that exiting from classical vacua is impossible. Consequently, the only viable path to particle production involves emergence from a quantum vacuum with zero scale factor $a=0$. Such an emergence can be interpreted as the Big Bang.

Crucially, the transition surface $a = 0, t = const$ (in conformal time: $a = 0,\quad  \eta \in (-\infty,+\infty)$ ) constitutes a light-like hypersurface rather than a space-like one, as commonly assumed. Particles are therefore produced on the light cone and are causally connected. The apparent simultaneity of particle creation is thus an illusion.

Since light propagation cannot be stopped, we conclude that the appropriate cosmological model must be open, i.e., $k=0$ or $k=-1$. Our model therefore describes an infinite Universe. In this framework, the Big Bang can be understood as a detonation wave propagating through the quantum vacuum at the speed of light—the only possible speed in a vacuum.

\roles{The authors contributed equally to this work. All authors have read and agreed to the published version of the manuscript.}

\funding{ The work of I.D. Ivanova was supported by Binational Science Foundation grants \#2021789 and \#2022132, by the ISF grant \#910/23 and by the Kreitman School of Advanced Graduate Studies of Ben-Gurion University of the Negev.}

\data{Data sharing is not applicable to this article as no new data were created or analyzed in this study.}

\end{document}